\documentclass[a4paper,12pt]{article}

\usepackage{amsthm,amsmath,amssymb,amsfonts}

\usepackage{graphicx}
\usepackage{caption}
\usepackage{subcaption}

\def\cR{{\cal R}}

\def\cH{{\cal H}}

\def\cJ{{\cal J}}
\def\s3{\sqrt{3}}
\def\s5{\sqrt{5}}

\newtheorem*{The*}{Theorem}
\newtheorem*{Rem*}{Remark}

\newcommand{\bear}{\begin{array}}
\newcommand{\eear}{\end{array}}

\newfont{\tenbi}{cmbxti10}
\newcommand{\beq}{\begin{equation}}
\newcommand{\eeq}{\end{equation}}

\newcommand{\rP}{\mathrm{P}}

\newcommand{\rV}{\mathrm{V}}

\newcommand{\rd}{\mathrm{d}}
\newcommand{\ri}{\mathrm{i}}

\newcommand\la{{\lambda}}

\newcommand\al{{\alpha}}
\newcommand\be{{\beta}}
\newcommand\gam{{\gamma}}

\begin{document}
\title{Generalizations of the short pulse equation}
\author{Andrew N.W. Hone\thanks{School of Mathematics, 
Statistics \& Actuarial Science, University of Kent, Canterbury, CT2 7NF, UK.}, 
Vladimir Novikov\thanks{Department of Mathematical Sciences, Loughborough University, 
Loughborough, LE11 3TU, UK.} and Jing Ping Wang$^*$
       }

\maketitle
\begin{abstract} 
We classify integrable scalar polynomial 
partial differential equations of second order generalizing the short pulse equation.  
\end{abstract} 

\section{Introduction}

The purpose of this short article is to present a classification of nonlinear partial differential equations of second order of the general form 
\begin{equation}
\label{eq}
u_{xt}=u+c_0u^2+c_1uu_x+c_2uu_{xx}+c_3u_x^2+d_0u^3+d_1u^2u_x+d_2u^2u_{xx}+d_3uu_x^2, 
\end{equation} 
which, in the case that   $c_j=0$ for $j=0,1,2,3$ and  $d_0=d_1=0$, $d_3=2d_2$, includes the short pulse equation
derived by Sch\"{a}fer and Wayne \cite{sw} as a model of ultra-short optical pulses in nonlinear media; cf. equation 
(\ref{eq2}) below. It was shown by 
Sakovich and Sakovich  \cite{saksp}  that the short pulse equation is integrable, in the sense that it admits a Lax pair and a recursion operator that generates infinitely many commuting symmetries;  these authors also found a hodograph-type transformation connecting it with the sine-Gordon equation. 
In fact, the short pulse equation  and the construction of its associated linear scattering problem first appeared in differential geometry \cite{beals, rabelo}. 
The integrability of the equation was further clarified by Brunelli \cite{brunelli}, who obtained a bi-Hamiltonian structure and used an alternative Lax representation to construct an infinite sequence of conserved quantities. 

The main result of this paper is the following. 


\begin{The*}
If the equation (\ref{eq}) 
possesses an infinite hierarchy of local higher symmetries, then up to rescaling 
$u\to\la u,\,x\to\mu x,\, t\to\nu t$ it is one of the list
\begin{eqnarray}
\label{eq1} u_{xt}&=&u+ ( u^2)_{xx},\\
\label{eq2} u_{xt}&=&u+(u^3)_{xx},\\
\label{eq3} u_{xt}&=&u+4uu_{xx}+u_x^2,  \\
\label{eq4} u_{xt}&=&u+(u^2-4u^2u_x)_{x},\\
\label{eq5} u_{xt}&=&u+2uu_{xx}+u_x^2,\\
\label{eq6} u_{xt}&=&u+u^2u_{xx}+uu_x^2,\\
\label{eq7} u_{xt}&=&u+\alpha(2uu_{xx}+u_x^2)+\beta( u^2u_{xx}+uu_x^2), \qquad \al\be\neq 0 .
\end{eqnarray}
\end{The*}

\begin{Rem*} The nonlinear terms in equation (\ref{eq7}) are a linear combination of those in equations (\ref{eq5}) and (\ref{eq6}). Upon applying an affine linear transformation $u\to au+b$ together with a Galilean transformation $x\to x-ct$ 
for suitable $a,b,c$,  the derivative of  (\ref{eq7}), that is 
$$ 
u_{xxt}= u_x+\alpha\Big(2uu_{xx}+u_x^2\Big)_x+\beta\Big( u^2u_{xx}+uu_x^2\Big)_x, 
$$ 
can be transformed to the case $\al=0$, $\be=1$, which is the derivative of (\ref{eq6}); but for the original 
equation  (\ref{eq7}) the quadratic terms cannot be removed in this way.
\end{Rem*} 

Equations of the form (\ref{eq}) are of interest for various reasons. Observe that, as written,  (\ref{eq}) is not an evolution equation for $u$, and if it is rewritten as one, solving for $u_t$, 
 then it becomes nonlocal, involving the  integration operator $D_x^{-1}$. 
  Physically, such equations appear in the description of  the 
short-wave behaviour of nonlinear systems. For example, the $b$-family of equations
\beq\label{bfam} 
m_t+um_x+bu_xm=0, \qquad m=m_0 +u-u_{xx} , \qquad  m_0=\mathrm{const},  
\eeq 
which was  introduced in \cite{dhh} 
(see also \cite{dhh2, hsta}) and derived from shallow water theory in \cite{dgh} and \cite{cl}, has a short-wave limit found by setting
$$ 
x\to \epsilon x, \qquad t\to\epsilon t, \qquad m_0\to -\mu\epsilon^{-2},\qquad  \mu =\mathrm{const},  
$$ 
and then taking $\epsilon \to 0$, 
which yields 
\beq\label{swhf} 
u_{xxt}+b\mu u_x + uu_{xxx}+bu_xu_{xx}=\Big(u_{xt}+b\mu u + uu_{xx}+\frac{1}{2}(b-1)u_x^2\Big)_x = 0. 
\eeq 
After sending $t\to -t$ and rescaling $u$, (\ref{swhf}) is seen to be the $x$ derivative of an equation of the form (\ref{eq}). It is known from \cite{miknov} that (for $b\neq 0$) the equation (\ref{bfam}) is integrable, 
in the sense that it admits an infinite hierarchy of commuting symmetries, if and only if $b=2$ (Camassa-Holm \cite{ch}) 
or $b=3$ (Degasperis-Procesi \cite{dp}).  Surprisingly, comparison with (\ref{eq1}), (\ref{eq3}) and (\ref{eq5}) in 
the above theorem shows that in the short-wave limit, there  are three integrable cases of equation (\ref{swhf}): 
not only  $b=2$ (Hunter-Saxton \cite{hs}) and $b=3$ (Vakhnenko \cite{vak}), but also the case $b=3/2$, which appears to be new. 


The proof of the above theorem consists of two parts. The first part consists of applying the perturbative symmetry approach, as described in \cite{miknov}, to obtain a set of necessary conditions on the parameters $c_j,d_j$ in (\ref{eq}) for the existence of a formal recursion operator with  local   coefficients (i.e. functions of $u$ and its derivatives only).  This part of the proof requires the use of computer algebra, and further details are omitted. Once a finite list of equations has been obtained as above (by scaling $c_j,d_j$ suitably), the remainder of the proof consists of explicitly constructing a recursion operator and associated infinite hierarchy of symmetries for each equation found. Thus in the rest of the paper we consider each equation on the list in turn, and for each one present the first higher symmetry, with flow variable $\tau$, and a recursion operator $\cal R$. The recursion operator  is factored as 
$\cR =\cH\cJ$, in terms of a compatible implectic-symplectic pair, with $\cH$ being a Hamiltonian operator such that the flow can be written as 
$$ 
u_\tau=\cH \, \delta_u \, \rho, 
$$ where $\rho$ is a density and $\delta_u$ denotes the variational derivative, i.e. 
$$
\delta_u \, \rho=\frac{\delta H}{\delta u} , \qquad \mathrm{where} \quad H=\int \rho \,\rd x 
$$ 
is the Hamiltonian functional with density $\rho$; and the flow is also written as 
$$
\cJ\, u_\tau = \delta_u \, \tilde{\rho}, 
 $$
using the symplectic operator $\cJ$ with 
 another density $ \tilde{\rho}$. In addition, for each item on the list we use a conservation law to define a reciprocal transformation, i.e. a change of independent variables of  hodograph type, which provides a link to other known integrable equations. We also present a Lax pair in each case.

Throughout the paper, subscripts with numbers are used to denote higher derivatives, so that $ u_{nx}=\frac{\partial^n u}{\partial x^n}$ for $n\geq 2$, 
but we also write e.g. $u_{xx}=u_{2x}$.

\section{Properties of the generalized short pulse equations} 

\subsection{Vakhnenko's  equation}

The equation (\ref{eq1}) was derived by Vakhnenko \cite{vak} as a model for the propagation of short-wave perturbations 
in a relaxing medium. Its loop soliton solutions were studied extensively in \cite{vp} and \cite{mpv}. In \cite{honewang} it 
was shown that the $x$ derivative of (\ref{eq1}) arises as a short-wave, high-frequency limit of the Degasperis-Procesi equation. Sometimes  (\ref{eq1})  is also referred to as the reduced Ostrovsky equation \cite{feng2, feng3}, 
since (up to rescaling dependent and independent variables) it is the special case $\be = 0$ of the Ostrovksy equation 
$$ 
\Big( u_t + uu_x + \be u_{xxx}\Big)_x -\gam u =0, 
$$ 
which is a model of weakly nonlinear ocean waves under the influence of the Coriolis force \cite{ost}. 

\noindent {\bf Higher symmetry:} 
The first higher symmetry of the equation (\ref{eq1}) is
\begin{equation}
 u_{\tau}=\left(\frac{u_{3x}}{(1+6u_{2x})^{\frac{5}{3}}}\right)_{xx}.\label{eq1s}
\end{equation}

\noindent {\bf Hamiltonian structure and recursion operator: } In terms of the quantity 
 $$w=(6 u_{xx}+1)^{-\frac{1}{3}},$$
the symmetry (\ref{eq1s}) becomes 
\beq\label{eq1w}
w_{\tau}=w^5w_{5x}+5 w^4 w_x w_{4x}+10 w^4 w_{2x}w_{3x}
\eeq 
which takes the form 
$$ w_{\tau}=-\frac{1}{2} \cH \delta_w w^{-1} , \qquad \mathrm{where} \quad 
\cH=w^4 D_x^5 w^4 $$
is a Hamiltonian operator. The associated symplectic operator is
\begin{equation}\label{sypw}
\cJ=w^{-2} D_x+D_x w^{-2} +(2 w^{-1} w_{2x}-w^{-2} w_x^2) D_x^{-1} 
w^{-2}+w^{-2} D_x^{-1} (2 w^{-1} w_{2x}-w^{-2} w_x^2) . 
\end{equation}
Thus the recursion operator $\cR=\cH \cJ$ generates the symmetries for 
(\ref{eq1}) and 
$$
\cJ w_{\tau}=\delta_w \left( -w^3 w_{3x}^2+\frac{8}{3} w^2 w_{2x}^3-4 
w w_x^2 w_{2x}^2-\frac{w_x^6}{3 w} \right) .
$$

\noindent {\bf Reciprocal transformation: }  
Viewed as a short-wave limit of the   Degasperis-Procesi equation, the $x$ derivative of  (\ref{eq1}) can be written in the 
form 
$$ 
m_t =2um_x+6u_xm, \qquad m=1+6u_{xx} ,
$$ 
giving a conservation law for the density $p=w^{-1}=(1+6u_{xx})^{1/3}$, that is 
\beq\label{hvak} 
p_t=2\, (up)_x, \qquad p^3=m. 
\eeq 
This conservation law leads to the introduction of new independent variables $X,T$ by means of 
the reciprocal transformation 
$$
\rd X=p \, \rd x+2pu\, \rd t, \qquad \rd T= \rd t, 
$$ 
so that (\ref{hvak}) produces
\beq\label{vakrt} 
(p^{-1})_T +2 u_X=0, \qquad p^3= 1+6p(pu_X)_X. 
\eeq 
If we use the letter $W$ to denote $u_x$, then we have
$$W=pu_X.$$ The equation (\ref{eq1}) becomes 
$$ 
W_T=u+2W^2, 
$$ 
and  (\ref{vakrt}) can be rewritten as the pair of relations 
$$ 
(\log p)_T = W, \qquad p^3-1 = 6pW_X, 
$$ 
which implies that $p$ satisfies the Tzitzeica equation in the form 
$$ 
(\log p)_{XT} = \frac{1}{6}\Big( p^2 - p^{-1}\Big). 
$$

\noindent {\bf Lax pair: } 
In \cite{pv}, a scalar Lax pair was presented for a reciprocally transformed version of  (\ref{eq1}), and in \cite{honewang} 
this was used to obtain a $3\times 3$ matrix Lax pair for the original equation, which is equivalent to the following 
 Lax representation with spectral parameter $\la$: 
\beq\label{1lax}
\mathbf{\Phi}_x=\left(\begin{array}{ccc} 0 & 1 & 0 \\ 
-2\lambda u_x & 0 & 1\\ 
-\frac{1}{3}\lambda & 2\lambda u_x & 0 \end{array}\right) \, \mathbf{\Phi}, \quad
\mathbf{\Phi}_t=\left(\begin{array}{ccc} 0& 2u & -\la^{-1} \\ 
 \frac{1}{3}(1-12\lambda u u_x)& 0 & 2u\\ 
\frac{4}{3}\la u& \frac{1}{3}(1+12\lambda u u_x)&0 \end{array}\right) \, \mathbf{\Phi}.
\eeq

\subsection{The short pulse equation}

The short pulse equation  was first derived as an equation for pseudospherical surfaces with an associated inverse scattering problem \cite{beals, rabelo}. Its physical derivation in nonlinear optics came later  \cite{sw}, and led to 
the construction of alternative forms of the Lax pair, recursion operator and bi-Hamiltonian structure \cite{brunelli, saksp}.

\noindent {\bf Higher symmetry:} 
The first higher symmetry of the equation (\ref{eq2}) is
$$
u_{\tau}=
\left(\frac{u_x}{(1+6u_x^2)^{\frac{1}{2}}}\right)_{xx}.
$$

\noindent {\bf Hamiltonian structure and recursion operator: } 
The above symmetry takes the Hamiltonian  form 
$$u_{\tau} =-\frac{1}{6}D_x \delta_u 
p
\qquad 
p= (6 u_x^2+1)^{\frac{1}{2}}, $$ 
with ${\cal H}=D_x$ being the Hamiltonian operator. 
The symmetries of equation (\ref{eq2}) are generated by the 
recursion operator
$$
\cR= D_x\cJ=
D_x \left( p^{-1} D_x p^{-1} +6 p^{-3} u_{2x} D_x^{-1} p^{-3} u_{2x} 
\right) , 
$$
and 
$$
\cJ u_{\tau}=
\delta_u \left(\frac{p^{-5}  u_{2x}^2}{2}\right) .
$$

\noindent {\bf Reciprocal transformation: }  
The equation  (\ref{eq2}) has the conservation law 
\beq\label{spcons} 
p_t=3\, (u^2 p)_x, 
\eeq 
which leads to the introduction of new independent variables $X,T$ according to 
$$ 
\rd X= p\, \rd x+ 3u^2 p \, \rd t, \qquad \rd T=\rd t. 
$$
In the new variables, the conservation law becomes 
$$
(p^{-1})_T+3\,(u^2)_X=0, 
$$ 
and by setting $W=u_x$ the original equation  (\ref{eq2})  gives 
\beq\label{wt} 
W_T=up^2, 
\eeq 
where we have used 
\beq\label{wprs} 
p^2 = 1+6W^2, \qquad u_X=\frac{W}{p}. 
\eeq 
Now if a new  dependent variable is introduced as 
$$ \varphi = -\frac{\ri}{2}\log\left(\frac{1+\ri\sqrt{6}W}{1-\ri\sqrt{6}W}\right) ,
$$ 
then from (\ref{wt}) and (\ref{wprs}) it follows that  
$$ 
u=\frac{\varphi_T}{\sqrt{6}}, 
$$ 
and $\varphi$ satisfies the 
sine-Gordon equation, that is 
$$ 
\varphi_{XT}=\sin\varphi . 
$$

\noindent {\bf Lax pair: } 
 Equation (\ref{eq2}) admits  the Lax representation 
\beq\label{2lax}
\mathbf{\Phi}_x=\left(\begin{array}{cc} 0 & 1+\ri\sqrt{6}u_{x}  \\ 
-\lambda(1-\ri\sqrt{6}u_x) & 0 \end{array}\right) \, \mathbf{\Phi},
\eeq
\beq\nonumber 
\mathbf{\Phi}_t=\left(\begin{array}{cc} {\ri\sqrt{6}u}/{2} &-\frac{1}{4\lambda}+3u^2+3\sqrt{6}\,\ri\, u^2u_{x}   \\ \frac{1}{4}+\lambda(-3u^2+3\sqrt{6}\,\ri \, u^2u_x)& - {\ri\sqrt{6}u}/{2} \end{array}\right) \, \mathbf{\Phi}.
\eeq 

\subsection{Equation (\ref{eq3}) } 

The equation  (\ref{eq3})  does not appear to have been considered before in the literature.

\noindent {\bf Higher symmetry:} 
The first higher symmetry of the equation (\ref{eq3}) is
\beq\label{eq3s}
u_{\tau}=\frac{u_{5x}}{(1+6u_{2x})^{\frac{10}{3}}}-30\frac{u_{3x}u_{4x}}{(1+6u_{2x})^{\frac{13}{3}}}+160\frac{u_{3x}^3}{(1+6u_{2x})^{\frac{16}{3}}}.
\eeq

\noindent {\bf Hamiltonian structure and recursion operator: } 
Let $w=(6 u_{2x}+1)^{-\frac{2}{3}}$. Then the symmetry (\ref{eq3s}) becomes 
\beq\label{eq3w}
w_{\tau}=w^5w_{5x}+5 w^4 w_x w_{4x}+\frac{5}{2} w^4 
w_{2x}w_{3x}+\frac{15}{4} w^3 w_x^2 w_{3x}=-2 \cH 
\delta_w w^{-1} ,
\eeq
where $$\cH=w^{\frac{5}{2}} D_x^2 w^{\frac{3}{2}} D_x  w^{\frac{3}{2}} D_x^2 
w^{\frac{5}{2}} $$ is a Hamiltonian operator. In terms of the quantity $w$, its symplectic operator has 
the same form as that for (\ref{eq1}), being  given by (\ref{sypw}).
Thus the recursion operator $\cR=\cH \cJ$ generates the symmetries for 
(\ref{eq3}) and 
$$
\cJ w_{\tau}=\delta_w \left( -w^3 w_{3x}^2+\frac{1}{6} w^2 w_{2x}^3-\frac{1}{4} 
w w_x^2 w_{2x}^2-\frac{w_x^6}{48 w} \right).
$$

\noindent {\bf Reciprocal transformation: } 
 After rescaling $u$ and taking $t\to-t$, the $x$ derivative of equation (\ref{eq3}) can be 
rewritten in the form 
\beq \label{beq} 
m_t+um_x+\frac{3}{2} u_x m=0, \qquad m= \frac{2}{3}+u_{xx}, 
\eeq
which is
a degenerate form of the $b$-family of peakon equations (\ref{bfam}), with $b=3/2$.
 The quantity 
$m^{2/3}$ is a conserved density, and the conservation law 
$$ 
p_t+(pu)_x=0, \qquad p=m^{2/3}
$$ 
 can be used to define the reciprocal transformation 
$$
\rd X=p \, \rd x-pu\, \rd t, \qquad \rd T= \rd t.
$$ 
Hence  (\ref{beq}) leads to the equations 
$$ 
(p^{-1})_T = u_X, \qquad p^{3/2}=\frac{2}{3}+p\Big(p u_X\Big)_X = \frac{2}{3}+p\Big(p(p^{-1})_T\Big)_X, 
$$ 
and the latter can be rewritten as 
\beq\label{tzit}
(\log p)_{XT}+p^{1/2}-\frac{2}{3}p^{-1}=0, 
\eeq 
which is equivalent to the Tzitzeica equation. 

\noindent {\bf Lax pair: }  Starting from a $3\times 3$ Lax  representation for the Tzitzeica equation, 
it is straightforward to obtain  the following Lax representation for (\ref{beq}):
\beq\label{3lax}
\mathbf{\Phi}_x=\left(\begin{array}{ccc} 0 & \mathrm{i}\lambda m & 0 \\ 
0 & 0 & \mathrm{i}\lambda m  \\ 
-\frac{2}{3}\lambda & 0 & 0 \end{array}\right) \, \mathbf{\Phi}, \qquad 
\mathbf{\Phi}_t=\left(\begin{array}{ccc} -\frac{1}{2}u_x & - \mathrm{i}\lambda um & \frac{1}{2}\la^{-1} \\ 
 \frac{\mathrm{i}}{2}\la^{-1} & 0 & - \mathrm{i}\lambda um \\ 
\frac{2}{3}\la u &  \frac{\mathrm{i}}{2}\la^{-1} & \frac{1}{2}u_x \end{array}\right) \, \mathbf{\Phi}.
\eeq

\subsection{Equation (\ref{eq4}) } 

To the best of our knowledge, the equation (\ref{eq4}) has not been studied before. 

\noindent {\bf Higher symmetry:} 
The first higher symmetry of the equation (\ref{eq4}) is
$$ 
u_{\tau}=\left(\frac{u_{4x}}{(1-2u_x)^{\frac{10}{3}}(1+4u_x)^{\frac{5}{3}}}
+10\frac{(8u_x-1)u_{2x}u_{3x}}{(1-2u_x)^{\frac{13}{3}}(1+4u_x)^{\frac{8}{3}}}+40\frac{(1-6u_x+24u_x^2)u_{2x}^3}{(1-2u_x)^{\frac{16}{3}}(1+4u_x)^{\frac{11}{3}}}\right)_x .
$$

\noindent {\bf Hamiltonian structure and recursion operator: } 
The above symmetry takes the form 
$$ u_{\tau}=\frac{1}{8}\cH \delta_u p, \qquad \mathrm{where} \quad 
\cH=D_x (1-2 u_x)^{-1}D_x (1-2 u_x)^{-1} D_x
$$
 is a Hamiltonian operator, 
and  $p$ is given by 
\beq\label{conp}
p=(1-2u_x)^{2/3}(1+4u_x)^{1/3} .
\eeq 
 The symmetry hierarchy
of (\ref{eq4}) can be generated by the recursion operator $\cR=\cH \cJ$, where $\cJ$ is a symplectic operator given by
\begin{eqnarray*}
 \cJ=D_x \left(fD_x+D_x f+gD_x^{-1}h+hD_x^{-1}g\right) D_x,
\end{eqnarray*}
with
\begin{eqnarray*}
 &&f=\frac{1}{2(1-2u_x)^{2}(1+4 u_x)^{2}}; \qquad g=\frac{8 u_x}{(1-2u_x)^{\frac{1}{3}}(1+4 u_x)^{\frac{2}{3}}};\\
 &&h=\frac{u_{3x}}{(1-2u_x)^{\frac{8}{3}}(1+4 u_x)^{\frac{7}{3}}}+\frac{2(10u_x-1)}{(1-2u_x)^{\frac{11}{3}}(1+4 u_x)^{\frac{10}{3}}}.
\end{eqnarray*}
Indeed, we have 
\begin{eqnarray*}
&&\cJ u_{\tau}=\delta_u 
\frac{1}{(4u_x+1)^{11/3}(1-2u_x)^{16/3}}\left(\frac{u_{4x}^2}{2}+\frac{2}{3}
\frac { (60 u_x-7)u_{3x}^2}{(4u_x+1)(2u_x-1)}\right.\\
&&\quad \left. -8 \frac{(360 u_x^2-62 u_x+17) 
u_{2x}^2 u_{3x}^2}{(4u_x+1)^2(2u_x-1)^2}+\frac{704}{15} \frac{(12960 u_x^4-4032 
u_x^3+2340 u_x^2-324 u_x+31) u_{2x}^6}{(4u_x+1)^4(2u_x-1)^4}\right).
\end{eqnarray*}

\noindent {\bf Reciprocal transformation: } 
The quantity $p$ 
in  (\ref{conp})
is a conserved density for (\ref{eq4}), with the conservation law 
\beq \label{5cons}
p_t = -4(u^2 p)_x, 
\eeq
leading to the reciprocal transformation 
\beq\label{trt} \rd X=p\, \rd x -4 u^2 p\, \rd t, \qquad \rd T = \rd t.
\eeq 
Under the latter change of independent variables, the conservation law (\ref{5cons}) 
 is transformed to the system 
\beq \label{rtsys}
(p^{-1})_T=4 (u^2)_X, \qquad 
\frac{(1-2W)^2(1+4W)}{p^3}=1, \quad W=pu_X ,
\eeq 
while the equation  (\ref{eq4})  becomes 
\beq\label{wpsi} 
W_T=\frac{up^2}{\psi}, 
\eeq 
where, from the second equation in (\ref{rtsys}), it is consistent to introduce the quantity $\psi$ 
such that 
\beq\label{psid}
\psi=\frac{1-2W}{p}, \qquad \frac{1}{\psi^2}=\frac{1+4W}{p}.
\eeq 
Then by (\ref{rtsys}) and (\ref{wpsi}) it follows that 
$$ 
(\log\psi)_T=-2u, 
$$ 
and by taking the $X$ derivative of the latter, using $u_X=W/p$, and taking the difference 
of the two equations in (\ref{psid}), an equation for $\psi$ alone results, namely 
\beq\label{tzi} 
(\log\psi)_{XT}=\frac{1}{3}\Big(\psi-\psi^{-2}\Big), 
\eeq 
which is a form of the Tzitzeica equation.

\noindent {\bf Lax pair: } 
The equation (\ref{eq4}) has the $3\times 3$  Lax representation 
\beq\label{4lax}
\mathbf{\Phi}_x=\left(\begin{array}{ccc} 0 & 1-2u_x & 0 \\ 
0 & 0 & 1+4u_x  \\ 
\lambda(1-2u_x) & 0 & 0 \end{array}\right) \, \mathbf{\Phi},
\eeq
\beq\nonumber 
\mathbf{\Phi}_t=\left(\begin{array}{ccc} 0& - 4u^2(1-2u_x)& \frac{1}{3}\la^{-1} \\ 
 \frac{1}{3}& 2u & - 4u^2(1+4u_x) \\ 
-4\la u^2(1-2u_x) &  \frac{1}{3} & -2u \end{array}\right) \, \mathbf{\Phi}.
\eeq 

It is interesting to apply the reciprocal transformation  (\ref{trt}) to the Lax pair. Upon making this 
change of independent variables, (\ref{4lax}) becomes 
\beq\label{rt4lax} 
\mathbf{\Phi}_X=\left(\begin{array}{ccc} 0 & p^{-1}-2u_X & 0 \\ 
0 & 0 & p^{-1}+4u_X \\ 
\lambda \, (p^{-1}-2u_X) & 0 & 0 \end{array}\right) \, \mathbf{\Phi}, \qquad
\mathbf{\Phi}_T=\left(\begin{array}{ccc} 0 &  0 & \frac{1}{3}\lambda^{-1} \\ 
 \frac{1}{3} & 2u & 0  \\ 
 0 &  \frac{1}{3} & -2u \end{array}\right) \, \mathbf{\Phi}.
\eeq 
 The compatibility of the linear system  (\ref{rt4lax}) 
 gives 
\beq \label{nrtsys}
(p^{-1})_T=4 (u^2)_X, \qquad 
u_{XT}-\frac{u}{p}-(u^2)_X=0 ,
\eeq 
where the second equation above arises as a differential consequence of the system (\ref{rtsys}), and can be 
integrated to yield the more general equation 
$$
\frac{(1-2W)^2(1+4W)}{p^3}
=1+F(X), \qquad W=pu_X ,
$$ 
where $F$ is an arbitrary function.  
However, upon making a point transformation in $X$, so that $$\hat{X}=G(X), \qquad u(X,T)=\hat{u}(\hat{X},T), 
\qquad p(X,T)=
G'(X)^{-1}\hat{p}(\hat{X},T),$$ 
the function $F$ can be removed by choosing  $G(X) =\int (1+F(X))^{1/3}\,\rd X$.

The system (\ref{rtsys}) corresponds to a negative flow in the Sawada-Kotera hierarchy. To see this, it 
is convenient to use the quantity $\psi$, as defined in (\ref{psid}),  
and then $\phi$,  the first component of the vector $\mathbf{\Phi}$, satisfies the scalar Lax pair 
\beq\label{sc4lax} 
\phi_{XXX}+V\,\phi_X=\lambda\, \phi, \qquad \phi_T=\frac{1}{3}\lambda^{-1}\Big(\psi\,\phi_{XX}-\psi_X\,\phi_X\Big), 
\eeq 
where $$ V=-\frac{\psi_{XX}}{\psi}.$$ If $V$ is not specified a priori, 
then the compatibility conditions for the scalar linear system are 
\beq\label{skminus1} 
V_T =-\psi_X, \qquad \psi_{XX}+V\,\psi =k, \qquad k_X=0, 
\eeq 
and, up sending to $T\to -T$, (\ref{sc4lax})  is equivalent to the  Lax pair found for the  
reciprocally transformed Vakhnenko equation in \cite{pv} (see \cite{honewang} for more details). In the case at hand, we have $k=0$, and substituting for $V$ in terms of $\psi$ in the first equation of  (\ref{skminus1}) and integrating produces 
$$ \psi^2\left((\log\psi)_{XT}-\frac{\psi}{3}\right)=\frac{\tilde{F}(T)}{3}, 
$$
where $\tilde{F}$ is an arbitrary function; after sending $\psi\to \tilde{F}(T)^{1/3}\,\psi$ and making a point transformation in $T$, this becomes the Tzitzeica equation in the form (\ref{tzi}).

\subsection{The Hunter-Saxton equation} 

In addition to the short-wave limit which takes the Camassa-Holm equation  (i.e. (\ref{bfam}) with $b=2$) to the equation  (\ref{eq5}), a further limit can be applied to remove the linear dispersion term. Taking the limit 
$$ 
x\to \epsilon x, \qquad t\to - \epsilon t,\qquad u\to \frac{1}{2} u, \qquad \epsilon \to 0
$$ 
produces the equation 
\beq\label{hseq} 
(u_t+uu_x)_x=\frac{1}{2}u_x^2; 
\eeq
a similar limit can be applied to remove the linear dispersion from  other  equations of the form (\ref{eq}). 
The equation (\ref{hseq}) was derived by Hunter and Saxton as an asymptotic model of liquid crystals \cite{hs}. The $x$ derivative of the Hunter-Saxton equation 
 corresponds to geodesic flow on an infinite-dimensional homogeneous space with constant positive curvature (see \cite{lenells} and references). 

\noindent {\bf Higher symmetry:} 
The first higher symmetry of the equation (\ref{eq5}) is
$$
u_{\tau}=\frac{u_{3x}}{(1+4u_{2x})^{\frac{3}{2}}}.
$$

\noindent {\bf Hamiltonian structure and recursion operator: } 
Notice that $$D_x u_{\tau}=-\frac{1}{4} \delta_u \sqrt{1+4 u_{2x}}.$$ 
Thus $D_x$ is a symplectic operator. The symmetries of (\ref{eq5}) can be generated by a recursion operator
\begin{eqnarray}\label{re5}
 \cR=\cH D_x=\left(\frac{1}{1+4 u_{2x}}D_x+D_x \frac{1}{1+4 u_{2x}}-8 u_{\tau} D_x^{-1} u_{\tau}\right) D_x .
\end{eqnarray}
The operators $\cH$ and $D_x^{-1}$ form a compatible Hamiltonian pair, which is a particular case of case V in Theorem 4 in \cite{wang10}.

\noindent {\bf Reciprocal transformation: } 
Considered as a short-wave limit of the Camassa-Holm equation, the $x$ derivative of  (\ref{eq5}) can be written in the 
form 
$$ 
m_t =2um_x+4u_xm, \qquad m=1+4u_{xx} ,
$$ 
giving the conservation law 
\beq\label{hsp} 
p_t=2\, (up)_x, \qquad p^2=m. 
\eeq 
Then introducing $X,T$ according to 
$$ \rd X = p\, \rd x + 2up\, \rd t , \qquad \rd T = \rd t, 
$$ 
and setting $W=u_x$, 
leads to the three equations 
\beq \label{pweqs} 
W_T=u+W^2, \qquad (p^{-1})_T +2u_X=0, \qquad p^2=1+4pW_X, 
\eeq 
where the first equation,  for $W_T$, comes from (\ref{eq5}), the second equation from 
(\ref{hsp}), and the third from the definition of $p$ in terms of $m$. Now from the second 
equation in (\ref{pweqs}) and the definition of $W$ it follows that 
$(\log p)_T = 2pu_X=2W$, so that upon differentiating the latter with respect to $X$ and using the 
third equation to eliminate $W_X$, an equation for $p$ alone results, namely 
$$ 
(\log p)_{XT}=\frac{1}{2}\Big( p -p^{-1}\Big).
$$ 
Thus, by setting $p=e^{\ri \varphi}$, this yields the sine-Gordon equation in the form $\varphi_{XT}=\sin\varphi$. 

\noindent {\bf Lax pair: } 
A Lax pair for the Hunter-Saxton equation in the form  (\ref{hseq}) was found in \cite{hz}. For  equation (\ref{eq5}), with the inclusion of linear dispersion, a  Lax representation  is
\beq\label{5lax}
\mathbf{\Phi}_x=\left(\begin{array}{cc} 0 & 1+4u_{xx}  \\ 
-\lambda & 0 \end{array}\right) \, \mathbf{\Phi},
\eeq
\beq\nonumber 
\mathbf{\Phi}_t=\left(\begin{array}{cc} u_x &-\frac{1}{4\lambda}+2u+8uu_{xx}   \\\frac{1}{4}-2\lambda u & -u_x \end{array}\right) \, \mathbf{\Phi}.
\eeq 

\subsection{The single-cycle pulse equation }

The equation (\ref{eq6}) was obtained recently by Sakovich \cite{sakovich} as a reduction of a coupled 
integrable short pulse system due to Feng \cite{feng}. Sakovich showed that the envelope soliton solution  of (\ref{eq6}) can only be as short as one cycle of its carrier frequency, and hence called it the single-cycle pulse equation.

\noindent {\bf Higher symmetry:} 
The first higher symmetry of the equation (\ref{eq6}) is
$$
u_{\tau}=\frac{u_{3x}}{(1+u_x^2)^3}-3\frac{u_xu_{2x}^2}{(1+u_x^2)^4}.
$$

\noindent {\bf Hamiltonian structure and recursion operator: } 
Notice that $$D_x u_{\tau}=\frac{1}{2} \delta_u \frac{u_{2x}^2}{(1+ u_{x}^2)^3}.$$ 
Thus $D_x$ is a symplectic operator. The symmetries of (\ref{eq6}) can be generated by a recursion operator
\begin{eqnarray*}
 \cR=\cH D_x=\left(\frac{1}{(1+ u_{x}^2)^2}D_x+D_x \frac{1}{(1+ u_{x}^2)^2}-4 u_xD_x^{-1}u_{\tau}-4  u_{\tau} D_x^{-1} u_x\right) D_x .
\end{eqnarray*}
The operators $\cH$ and $D_x^{-1}$ form a compatible Hamiltonian pair, which is a particular case of case IV in Theorem 4 in \cite{wang10}.

\noindent {\bf Reciprocal transformation: } 
From the conservation law 
$$ 
p_t=(u^2\, p)_x, \qquad p =1+u_x^2, 
$$ 
the reciprocal transformation 
\beq\label{RT6} 
\rd X = p\, \rd x + u^2 p \, \rd t, \qquad \rd T = \rd t 
\eeq 
yields  the equations 
\beq \label{eq6rt} 
(p^{-1})_T+(u^2)_X=0, \qquad p=1+p^2u_X^2. 
\eeq 
To see how the latter system is related to the sine-Gordon equation, it is most convenient to consider the Lax pair 
\beq\label{rt6lax}
\mathbf{\Phi}_X=\left(\begin{array}{cc} -\ri A & 1  \\ 
-\lambda  & \ri A \end{array}\right) \, \mathbf{\Phi},
\qquad 
\mathbf{\Phi}_T=\left(\begin{array}{cc} 0  &   B\,\la^{-1} \\ -\overline{B}  
 &  0  \end{array}\right) \, \mathbf{\Phi} 
\eeq  
where 
$$ 
A = u_{XX}+p^{-1}p_Xu_X, \quad  B = \frac{1}{4}\left(\frac{pu_X+\ri}{pu_X-\ri}\right), \quad 
\overline{B} = \frac{1}{4}\left(\frac{pu_X-\ri}{pu_X+\ri}\right). 
$$ 
The compatibility conditions for this Lax pair mean that it is consistent to set 
$$ 
A=-\frac{1}{2}\vartheta_X, \quad B=\frac{1}{4}e^{\ri\vartheta}, \quad \overline{B}=\frac{1}{4}e^{-\ri\vartheta}, 
$$
where $\vartheta$ satisfies 
$$ 
\vartheta_{XT} +\sin \vartheta =0.
$$ 
The solution of (\ref{eq6rt}) is given in terms of the variable $\vartheta$ by 
$$ 
u=-\frac{1}{2}\vartheta_T, \qquad p=\mathrm{sec}^2(\vartheta/2). 
$$

\noindent {\bf Lax pair: }  
Using the inverse of the reciprocal transformation  (\ref{RT6}) 
to rewrite the Lax pair (\ref{rt6lax})  in terms of the original independent variables $x,t$ 
gives 
a Lax representation for equation (\ref{eq6}), namely 
\beq\label{6lax}
\mathbf{\Phi}_x=\left(\begin{array}{cc} -\frac{\ri u_{xx}}{1+u_x^2} & 1+u_x^2  \\ 
-\lambda(1+u_x^2) & \frac{\ri u_{xx}}{1+u_x^2} \end{array}\right) \, \mathbf{\Phi},
\eeq 
\beq\nonumber
\mathbf{\Phi}_t=\left(\begin{array}{cc} -\frac{\ri u^2u_{xx}}{1+u_x^2} &
\frac{(u_x+\ri)}{4\lambda(u_x-\ri )}  +u^2(1+u_x^2) 
 \\-  \frac{(u_x-\ri)}{4 (u_x+\ri )}   -\la  u^2(1+u_x^2) 
 & \frac{\ri u^2 u_{xx}}{1+u_x^2} \end{array}\right) \, \mathbf{\Phi}.
\eeq

\subsection{Equation (\ref{eq7}) } 

As noted in the remark above, the equation  (\ref{eq7}) combines the nonlinear terms from 
 (\ref{eq5}) and  (\ref{eq6}), but cannot be directly reduced to either equation.

\noindent {\bf Higher symmetry:} 
The first higher symmetry of the equation (\ref{eq7}) is
$$
u_{\tau}=\frac{u_{3x}+3\frac{\beta}{\alpha}(1+\beta u_x^2)\, u_xu_{2x}+\frac{\beta}{2\alpha^2}(1+\beta u_x^2)^3\, u_x}{((1+\beta u_x^2)^2+4\alpha u_{2x})^{\frac{3}{2}}} .
$$
This reduces to the first higher symmetry of 
 (\ref{eq5})  when $\beta=0$ and $\alpha=1$, but does not behave well in the limit $\al\to 0$. 

\noindent {\bf Hamiltonian structure and recursion operator: } 
Similarly to the previous case we have $$D_x u_{\tau}=-\frac{1}{4\alpha^2} \delta_u\left((1+\beta u_x^2)^2+4 \alpha u_{2x}\right)^{\frac{1}{2}}.$$
A recursion operator is given by
\begin{eqnarray*}
 \cR=\left(\frac{1}{(1+\beta u_x^2)^2+4 \alpha u_{2x}}D_x+D_x \frac{1}{(1+\beta u_x^2)^2+4 \alpha u_{2x}}-8 \alpha^2 u_{\tau} D_x^{-1}u_{\tau}+\frac{2\beta^2}{\alpha^2} u_{x} D_x^{-1} u_x\right) D_x .
\end{eqnarray*}
Notice that when $\beta=0$ and $\alpha=1$, it leads to the recursion operator (\ref{re5}).

\noindent {\bf Reciprocal transformation: } 
The equation (\ref{eq7})  can be rewritten as 
\beq\label{vform}
u_{xt}=u+vu_{xx}+\frac{1}{2}v_xu_{x}, \qquad v=2\al u+\be u^2,  
\eeq
which  leads to an equation with a form analogous to the Camassa-Holm equation \cite{ch}, 
namely  
$$ 
m_t = vm_x + 2v_x m, 
$$ 
where  
\beq\label{mdef}
 m = (1+\be u_x^2)^2 +4\al u_{xx}, 
\eeq 
and this gives the conservation law 
\beq\label{eq7cons} 
p_t=(vp)_x, \qquad p= m^{1/2}. 
\eeq 
Upon introducing the reciprocal transformation 
$$ 
\rd X = p\, \rd x + pv\, \rd t, \qquad \rd T = \rd t, 
$$ 
the above relations between $u$ and $p$ are transformed to 
\beq\label{rteq7} 
(p^{-1})_T + 2(\al + \be u)u_X=0, \qquad p^2 = (1+\be p^2u_X^2)^2+4\al p(pu_X)_X. 
\eeq 

In order to identify this symmetry in terms of known integrable equations of third order, we extend the above reciprocal transformation to include this additional flow. First of all, note that 
$$ 
u_\tau = \frac{1}{2\al}\left(\frac{p_x}{p^2}+\frac{\be u_x\rP}{\al p}\right), \qquad
\rP=1+\be u_x^2, 
$$
which allows us to   write  
$$ 
\rd X = p\, \rd x + pv\, \rd t +F\, \rd \tau, \qquad \rd T = \rd t, \qquad \rd S=\rd \tau,  
$$ 
where 
$$ F=\frac{p_{xx}}{p^3}-\frac{3p_x^2}{2p^4}+\frac{\be\rP}{4\al^2}\Big(3-\frac{\rP^2}{p^2}\Big). 
$$ 
By applying the above reciprocal transformation to the symmetry, we consider the ratio $\rho = p/\rP$, 
and then 
set 
$$ 
\rho = \frac{p}{\rP}=e^{\ri\vartheta} 
$$ to find the equation 
\beq\label{CDeq} 
\vartheta_S=\vartheta_{XXX}+\frac{1}{2}\vartheta_X^3 -\frac{3\be}{2\al^2}\,\vartheta_X\sin^2 \vartheta , 
\eeq 
which is a form of the Calogero-Degasperis-Fokas equation (see \cite{cd,fokas,hlav}).
The equation (\ref{CDeq}) is related via the Miura transformation 
\beq\label{Vmiura} 
y=-\ri\, \frac{\rho_X}{\rho} -\frac{\sqrt{\be}}{2\al}\Big( \rho- \rho^{-1}\Big)
=\vartheta_X-\ri\frac{\sqrt{\be}}{\al}\sin \vartheta 
\eeq 
to the modified KdV (mKdV) equation in the form 
\beq\label{mkdv} 
y_S=y_{XXX}+\frac{3}{2} y^2y_X.
\eeq 
Thus we see that under a reciprocal transformation, the equation (\ref{eq7}) corresponds to a symmetry of the mKdV 
equation. 

The calculations involving the reciprocal transformation are most conveniently carried out by introducing the variable $W=u_x$, so that (\ref{eq7}) becomes 
$$ 
W_T =u \rP+\al W^2, \qquad \mathrm{with}\quad  W=pu_X, \quad \rP=1+\be W^2, 
$$ 
and the second equation in (\ref{rteq7}) gives 
\beq\label{wx} 
W_X = \frac{p^2-\rP^2}{4\al p} \implies \frac{W_X}{\rP}=\frac{\ri}{2\al}\, \sin\vartheta. 
\eeq 
Then for $\rho = p/\rP$ we find 
\beq\label{wp} 
(\log\rho)_T=\frac{2\al W}{\rP}  \implies \frac{W}{\rP}=\frac{\ri}{2\al}\, \vartheta_T.  
\eeq 
Upon computing the  $X$ derivative of both sides, this yields 
$$ 
(\log\rho)_{XT}+\frac{1}{2}\Big( \rho - \rho^{-1}\Big) = \frac{1}{\rP}\Big( \rho - \rho^{-1}\Big) 
$$ 
or equivalently 
$$ 
 \frac{2}{\rP} = 
\frac{\vartheta_{XT}}{\sin\vartheta}+1,  
 $$ 
which indicates that the symmetry of the Calogero-Degasperis-Fokas equation corresponding to (\ref{eq7}) is not the sine-Gordon equation, but 
something of higher order. Indeed, differentiating the above equation with respect to $X$ and using 
$\rP_X=2\be WW_X$ together with (\ref{wx}) and (\ref{wp}) leads to the third order equation 
\beq\label{sggen} 
  \left(\frac{\vartheta_{XT}}{\sin\vartheta}\right)_X+ \frac{\be}{\al^2}\Big(\cos\vartheta\Big)_T =0. 
\eeq 
In terms of these transformed coordinates, the Lax pair for the $T$ flow takes the 
form 
\beq\label{rt7lax}
\mathbf{\Phi}_X=\left(\begin{array}{cc} \frac{\ri y}{2}  & 1  \\ 
-\lambda  & -\frac{\ri y}{2}  \end{array}\right) \, \mathbf{\Phi},
\qquad 
\mathbf{\Phi}_T=\left(\begin{array}{cc} 0 &   \eta \,\la^{-1} \\ \zeta
 &  0  \end{array}\right) \, \mathbf{\Phi} ,
\eeq  
where $y$ is given by (\ref{Vmiura}) and 
\beq\label{eta} 
\eta=-\frac{e^{\ri\theta}}{4}\left( \frac{\theta_{XT}} {\sin\theta} +\frac{\sqrt{\be}}{2\al} \theta_T \right) , 
\qquad 
\zeta= -\frac{\ri}{2}y_T-\eta. 
\eeq 
With the introduction of  the KdV potential 
\beq\label{kdvpot}
\rV= -\frac{\ri}{2}y_X+\frac{1}{4}y^2, 
\eeq 
this corresponds  to the  negative KdV flow (see \cite{honewang}) given by 
\beq\label{negkdv}
\rV_T=2\eta_X, \qquad \eta_{XXX}+4\rV\eta_X + 2\rV_X\eta=0, 
\eeq 
but not the general solution of this. Indeed, integration of the second equation in (\ref{negkdv}) 
gives  
$$ 
\eta\eta_{XX}-\frac{1}{2}\eta_X^2+2\rV\eta^2 = F(T), 
$$ 
(a form of the Ermakov-Pinney equation, cf. equation (4) in  \cite{cm}) 
but from the expression (\ref{eta}) and (\ref{sggen}) it follows that 
$$ 
F(T)=0, \qquad y = -\ri (\log\eta)_X, \qquad \rV=-\frac{\eta_{XX}}{2\eta}+\frac{\eta_X^2}{4\eta^2}, 
$$ 
in accordance with the compatibility of the linear system (\ref{rt7lax}). 
Substituting the latter expression for $\rV$ in terms of $\eta$ into the first equation in (\ref{negkdv}) yields an equation 
of third order for $\eta$, which integrates to yield 
$$ 
\eta(\log \eta)_{XT}+2\eta^2 = 2G(T), 
$$ 
and, after rescaling $\eta\to \sqrt{G(T)}\, \eta$ and redefining $T$ so that $\partial_T \to \sqrt{G(T)}\, \partial_T$, 
we see that $\varphi=\ri \log\eta$ satisfies the sine-Gordon equation in the form  $\varphi_{XT}+4\sin\varphi = 0$. 

\noindent {\bf Lax pair: }  
In order to obtain a Lax pair for the equation (\ref{eq7}), it is sufficient to rewrite (\ref{rt7lax}) in terms of the original 
independent variables $x,t$. However, due to the dependence on $p$, this does not directly produce matrices which are 
rational functions of the original field $u$ and its derivatives. To obtain a rational Lax pair, it is convenient to put (\ref{rt7lax}) into scalar form and carry out a gauge transformation, 
which leads to the scalar linear system 
\beq\label{7scalax}
\psi_{xx}+\left(m\la + \frac{f_{xx}}{f}-2\, \frac{f_x^2}{f^2}\right)\, \psi=0, \qquad 
\psi_t = \left(v - \frac{1}{4f^2\la}\right) \psi_x -\left(\frac{1}{2}v_x+\frac{f_x}{4f^3\la}\right)\,\psi, 
\eeq 
where $m$ is given in (\ref{mdef}), $v$ is as in (\ref{vform}), and 
$$ 
f=1+\ri\sqrt{\be}\, u_x. 
$$ 
The scalar system (\ref{7scalax}) can readily be put into matrix form if desired.

\section{Conclusions} 
The list of integrable generalized  short pulse equations appears to contain three new equations, namely  (\ref{eq3}),  (\ref{eq4}), and  also (\ref{eq7}), which combines the nonlinear terms of the Hunter-Saxton equation and the single-cycle pulse equation. All of the equations considered here are related by a reciprocal transformation to either the sine-Gordon equation or the Tzitzeica equation. However, in the case of equation  (\ref{eq7}), the link is rather indirect, and the equation that arises directly is the symmetry (\ref{sggen}) of the Calogero-Degasperis-Fokas equation, which does not seem to have been considered before. These reciprocal links should be examined further, in order to derive explicit solutions of  the new equations in parametric form. 

\noindent 
{\bf Acknowledgments:} ANWH is supported by Fellowship EP/M004333/1 from the Engineering and Physical Sciences 
Research Council (EPSRC). 


\begin{thebibliography}{99} 

\bibitem{beals} R. Beals, M. Rabelo and K. Tenenblat,  
Stud. Appl. Math. {\bf 81} (1989) 125--151.

\bibitem{brunelli} J.C. Brunelli, Phys. Lett. A {\bf 353} (2006) 475--478. 

\bibitem{ch} R. Camassa and D.D. Holm, 
Phys. Rev. Lett. {\bf 71}  (1993) 1661--1664.

\bibitem{cd} F. Calogero and A. Degasperis,  J. Math. Phys. {\bf 22} (1981) 23.

\bibitem{cm}  A.K. Common and M. Musette, Phys. Lett. A {\bf 235} (1997) 574--580. 

\bibitem{cl} 
A. Constantin and D. Lannes, Archive for Rational Mechanics and Analysis
{\bf 192} (2009) 165--186.

\bibitem{dp}  A. Degasperis and M. Procesi, 
Asymptotic integrability.
Symmetry and Perturbation Theory,  eds. A. Degasperis and G. Gaeta. World Scientific (1999) pp. 23--37.

\bibitem{dhh} A. Degasperis, D.D. Holm and A.N.W. Hone, Theor. Math. Phys. {\bf 133} (2002)  1463--1474.

\bibitem{dhh2} A. Degasperis, D.D. Holm and A.N.W. Hone, Integrable and non-integrable equations with peakons, 
Proceedings of Nonlinear Physics - Theory and Experiment II, World Scientific (2002) 37--43.

\bibitem{dgh} H.R. Dullin, G.A. Gottwald and D.D. Holm, Fluid Dynamics Research
{\bf 33 } (2003) 73--95.

\bibitem{feng} B.-F. Feng, 
J. Phys. A: Math. Theor. {\bf 45} (2012) 085202.

\bibitem{feng2} B.-F. Feng, K. Maruno and Y. Ohta, 
J. Phys. A: Math. Theor. {\bf 45} (2012) 355203. 

\bibitem{feng3} B.-F. Feng, K. Maruno and Y. Ohta, 
J. Phys. A: Math. Theor. {\bf 48} (2015) 135203. 

\bibitem{fokas} A.S. Fokas, J. Math. Phys. {\bf 21} (1980) 1318. 

\bibitem{hlav} L. Hlavaty, Phys. Lett. A {\bf 113} (1985) 177--178. 

\bibitem{hs} D.D. Holm and  M. Staley, 
Phys. Lett. A {\bf 308} (2003)  437--444. 



\bibitem{honewang} A.N.W. Hone and J.P. Wang, 
Inverse Problems {\bf 19 } (2003) 129--145. 

\bibitem{hsta} J.K. Hunter, R. Saxton, 
SIAM J. Appl. Math. {\bf 51} (1991) 1498--1521.

\bibitem{hz} J.K. Hunter and Y. Zheng, 
Physica D {\bf 79} (1994) 361--386. 

\bibitem{lenells}J. Lenells, 
J. Geom.  Phys. {\bf  57 } (2007) 2049--2064. 

\bibitem{miknov}  A.V.  Mikhailov and  V.S. Novikov, 
J. Phys. A: Math. Gen. {\bf 35} 
(2002)
4775--4790.

\bibitem{mpv} A.J. Morrison,  E.J. Parkes,  and  V.O. Vakhnenko, 
Nonlinearity {\bf 12} (1999) 1427--1437. 

 \bibitem{ost} L.A. Ostrovsky, 
Oceanology
{\bf 18} (1978)
119--125.

\bibitem{pv} E.J. Parkes  and  V.O. Vakhnenko,  Chaos Solitons Fractals {\bf 13} (2002) 1819--26.


\bibitem{rabelo} 
M.L. Rabelo, 
Stud. Appl. Math. {\bf 81} 
(1989) 221--248. 

\bibitem{saksp} A. Sakovich and S. Sakovich, J. Phys. Soc. Jpn. {\bf 74}  (2005) 239. 

\bibitem{sakovich}S. Sakovich, 
Commun. Nonlinear Sci. Numer. Simulat. {\bf 39 } (2016) 21--28. 

\bibitem{sw} T. Sch\"{a}fer and  C.E. Wayne, 
Physica D {\bf 196} (2004) 90--105.
 
\bibitem{vak} V.O.  Vakhnenko, 
J. Phys. A: Math. Gen.{\bf 25} (1992)  4181--4187. 

\bibitem{vp}V.O. Vakhnenko and E.J. Parkes, 
Nonlinearity {\bf 11} (1998)  1457--1464. 



\bibitem{wang10} J.~P.~Wang,
Nonlinearity 
 {\bf 23} 
(2010) 2009--2028.

\end{thebibliography}
\end{document}